\documentclass[doublecolumn]{elsart}

\usepackage{natbib, graphicx}

\usepackage{amssymb}
\journal{Physica A}
\begin{document}

\begin{frontmatter}

\title{On high energy tails in inelastic gases}


\author[brenig,lambiii]{R. Lambiotte},
\corauth[cor]{Corresponding author.} 
\ead{rlambiot@ulb.ac.be}
\author[brenig]{L. Brenig} and
\ead{lbrenig@ulb.ac.be}
\author[marcos]{J.M. Salazar}
\ead{jmarcos@u-bourgogne.fr}

\address[brenig]{Physique Statistique, Plasmas et Optique Non-lin\'eaire, Universit\'e Libre de Bruxelles, Campus Plaine, Boulevard du Triomphe, Code Postal 231, 1050 Bruxelles, Belgium}
\address[lambiii]{SUPRATECS, Universit\'e de Li\`ege, B5 Sart-Tilman, B-4000 Li\`ege, Belgium}
\address[marcos]{LRRS UMR-5613 CNRS,  Facult\'e des Sciences Mirande, 9 Av. Alain Savary 21078 Dijon Cedex , France.}

\begin{abstract}
We study the formation of high energy tails in a one-dimensional  kinetic model for granular gases, the so-called Inelastic Maxwell Model. We introduce a time-discretized version of the stochastic process, and show that  continuous time implies larger fluctuations of the particles energies. This is due to a statistical relation between the number of inelastic collisions undergone by a particle and its average energy. This feature is responsible for the high energy tails in the model, as shown by computer simulations and by analytical calculations on a linear Lorentz model.
\end{abstract}

\begin{keyword}
Granular models of complex systems \sep Random walks and L\'evy flights \sep Kinetic theory
\PACS 45.70.Vn\sep 05.40.Fb\sep 45.20.Dd

\end{keyword}

\end{frontmatter}

\section{Introduction}

Lorentz gas models have been used historically in order to clarify the features of more complex non-linear kinetic equations. For instance, P. and T. Ehrenfest \cite{ehrenfest} introduced the so-called wind tree model, in order to discuss the {\em Stosszahlansatz} of the Boltzmann equation. In this paper, we will use  a similar approach in order to study inelastic gases, namely low-density systems composed by a large number of moving macroscopic particles, themselves performing inelastic interactions. 
Due to their inelasticity, inelastic gases dissipate kinetic energy. This implies that  the granular temperature, defined kinetically by T$\sim$$ <$$V^{2}$$>$, where  ${\bf V}\equiv {\bf v} - {\bf u}(r, t) $ and {\bf u}(r, t) is the local mean velocity, asymptotically vanishes if the system is not supplied by an external energy source. Before reaching total rest state, these systems usually reach 
a  self-similar solution, i.e. form preserving solution whose time dependence occurs through the granular temperature,
$f(v; t) = \frac{1}{\sqrt{T(t)}} f_S(\frac{v}{\sqrt{T(t)}})$.
Such scaling velocity distributions have been observed in a large variety of kinetic models, and have been shown to generically highlight non-Maxwellian features and  overpopulated high energy tail \cite{bennaim1,ernst}.
Let us note that the specific shape of the tail may exhibit very different behaviours, depending on the details of the model  \cite{biben}. 

In this paper, we show on simplified kinetic models that such high energy tails may originate from the fact that the average energy of particles depends on their collision history, i.e. on the number of collisions they have undergone in the course of time. This effect, that has no counterpart in the case of elastic interactions, is verified by Direct Simulation Monte-Carlo (DSMC) simulations of the non-linear Inelastic Maxwell Models, and by focusing on a discrete time version of the dynamics. Analytical calculations are also performed for a simpler linear 
Lorentz model.

\section{Inelastic Maxwell Model}
In the following, we study scaling solutions in the context of the one-dimensional Inelastic Maxwell Model (IMM). This model \cite{bennaim1} derives from a mathematical simplification of the non-linear Boltzmann equation for inelastic hard rods \cite{barrat}, assuming that the collision rate may be replaced by an average rate, independent of the relative velocity of the colliding particles. We also assume that the system is and remain homogeneous in the course of time, so that the resulting kinetic equation writes:

\begin{equation}
 \frac{\partial f(v_{1};t)  }{\partial t}  +f(v_{1};t) = \int_{-\infty}^{\infty}  dv_{2} ~   (\frac{1}{\alpha}) 
f(v^{'}_{1};t) f(v^{'}_2;t)   
\label{imm}
\end{equation} 
where the primed velocities are the pre-collisional velocities,  defined by the collision rule:

\begin{equation}
v_{1}^{'} = v_{1} - \frac{(1+ \alpha )}{2 \alpha} v_{12}  ~~~~
v_{2}^{'} = v_{2} + \frac{(1+ \alpha )}{2 \alpha} v_{12}
\label{collision}
\end{equation} 
$v_{12}$ is the relative velocity $v_1 - v_2$. The restitution parameter
$\alpha$ $\in ~ ]0:1]$ measures dissipation of energy at each collision, and the elastic limit corresponds to the case $\alpha=1$.
It is well-known that the scaling  solution of \ref{imm} reads \cite{balda1}:

\begin{equation}
f(v; t) = \frac{1}{\pi \sqrt{T(t)}} \frac{1}{[1 + (\frac{v}{\sqrt{T(t)}})^{2}]^{2}}
\label{baldassari}
\end{equation}
where the granular temperature decreases like $T(t) = T_0 ~ e^{-\frac{(1-\alpha)^2 t}{2}}$.

Recently, we have introduced \cite{lambi} a time discretized version of \ref{imm}:

\begin{equation}
f_{N+1}(v_{1}) =  \int_{-\infty}^{\infty}  dv_{2} ~   (\frac{1}{\alpha}) 
f_N(v^{'}_{1}) f_N(v^{'}_2)   
\label{discrete}
\end{equation} 
Let us call continuous time dynamics (CTD) and discrete time dynamics (DTD) the models associated to \ref{imm} and \ref{discrete} respectively.
Arguments based on the central limit theorem show that any initial velocity distribution converges towards a scaling solution whose shape is gaussian in the case of DTD:

\begin{figure}

\includegraphics[angle=-90,width=4.0in]{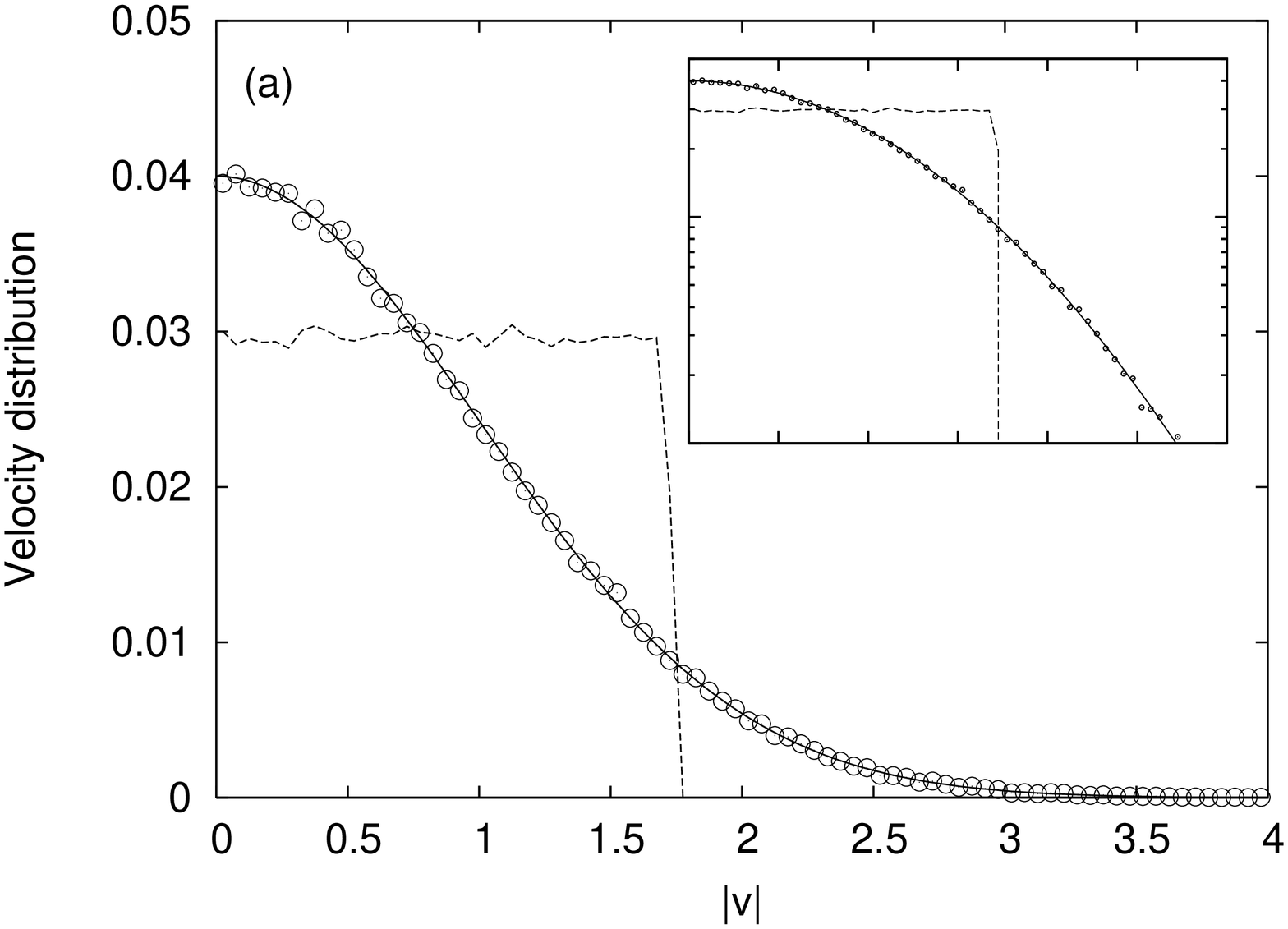}

\includegraphics[angle=-90,width=4.0in]{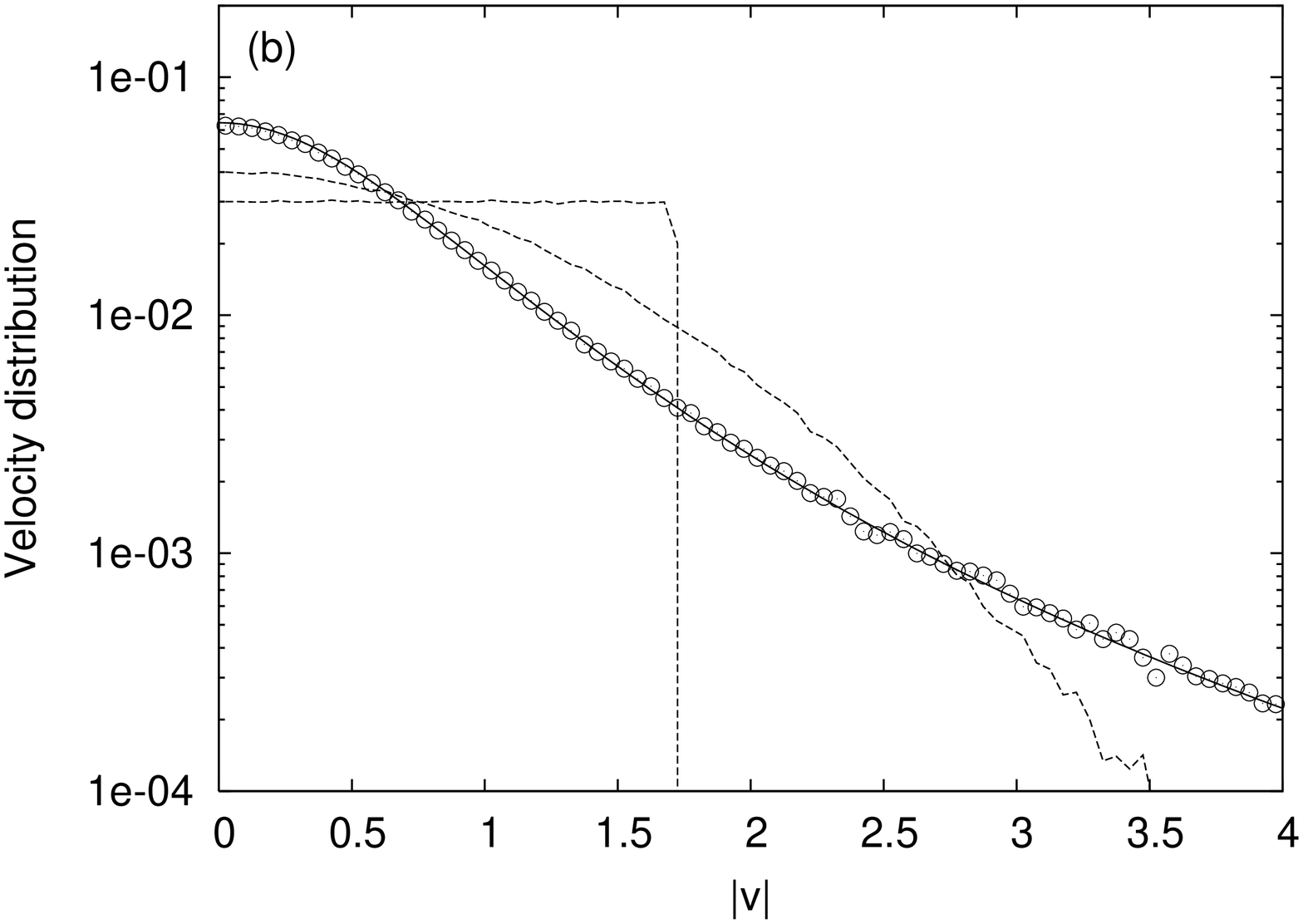}
\caption{\label{figure1}  In (a), velocity distributions in arithmetic and in logarithmic scale (inset) for DTD. The solid line corresponds to the theoretical Maxwell-Boltzmann distribution. In (b), the simulations of CDT show the convergence towards  the theoretical solution \ref{baldassari} (solid line). In both cases, the system is composed by 150000 particles with $\alpha$=0.5.
 The asymptotic velocity distribution is plotted with circles ($\circ$). The dashed lines are guides for the eye in order to isolate the initial conditions, that are uniform or Maxwell-Boltzmann distribution.
}
\end{figure}

\begin{figure}

\includegraphics[angle=-90,width=4.0in]{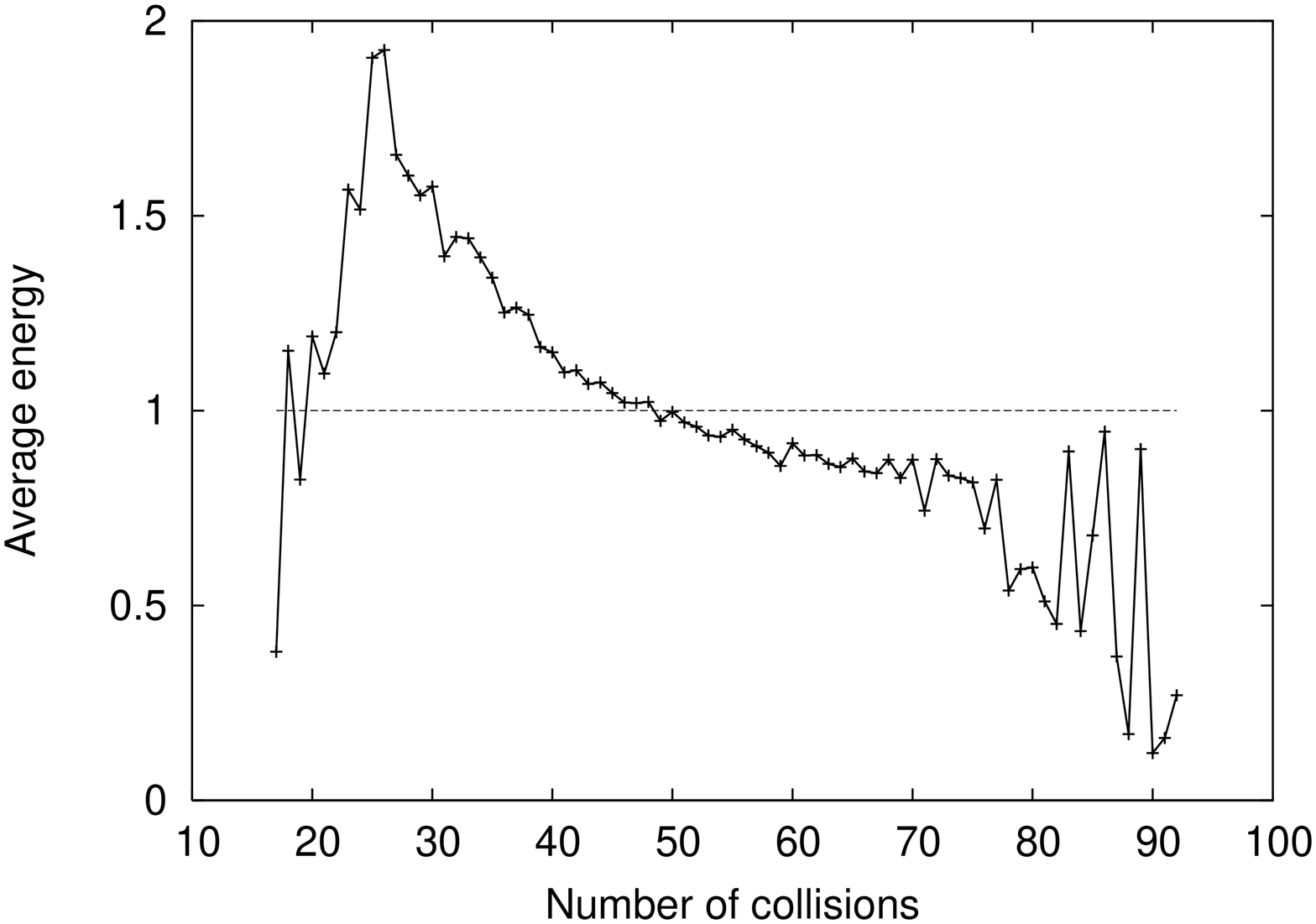}

\caption{\label{figure2} Average energy of particles as a function of their number of collisions. The dashed line is the constant value $<E>=1$. The system is composed of 20000000 particles with $\alpha=0.5$. 
}
\end{figure}

\begin{equation}
f_{N}(v) = \frac{1}{(2 \pi T_N)^{\frac{1}{2}}} e^{- \frac{v^2}{2 T_N}}
\label{maxwell}
\end{equation} 
$T_N$ is the temperature after N collision steps, $T_N=(\frac{1+\alpha^2}{2})^N T_0$.  One should stress that the transition from a power law tail \ref{baldassari} towards a Maxwell-Boltzmann distribution \ref{maxwell} is entirely due to the passage from a continuous time towards a discrete time description. 
The main difference between these two models follows. In DTD, every particle  has suffered exactly $N$ inelastic collisions at time $N$, so that it has dissipated the same fraction of energy on the average at that time. In contrast, in CDT, the constituents of the inelastic gas have performed different number of inelastic collisions $N$ at a given time t, distributed according to the Poisson law $\frac{t^N e^{-t}}{N!}$. This suggests that ensembles of particles, discriminated by their  number of collisions, should be characterised by different quantities of energy on average, thereby increasing fluctuations of the particles energies.

In order to verify this possible explanation, we have performed DSMC simulations \cite{bird} of the kinetic equations \ref{imm}
 and \ref{discrete}, based on their interpretation as a stochastic process.
 The CDT case is well-known
and consists in picking a given number of random collision pairs at each step.  The DTD case  is based on the same algorithm, except that at each step, the N particles are distributed randomly into $\frac{N}{2}$ collision couples. This method ensures that each particle collides one and only one time at each step. At the end of each step, the velocities are rescaled in order to keep the total energy constant \cite{soto}. Simulation methods have been verified by looking at the asymptotic velocity distribution, and checking the theoretical results \ref{baldassari} and \ref{maxwell} (fig.\ref{figure1}). In the case of CDT, we have also  highlighted a non-trivial relation between the number of collisions suffered by a particle, and its average energy. In order to evaluate this relation, we have started a simulation from a Maxwell-Boltzmann initial condition, and let the simulation run during 50 collisions per particle. After that time, the simulation is stopped, the particles velocities are rescaled so that the average energy $<E>=1$ and we measure the average energy $<E>_K$ of particles having performed $K$ collisions. The results (fig.\ref{figure2}) clearly show that the average energy of particles is a decreasing function of their number of collisions, thereby confirming the above discussion.

\section{One dimensional Lorentz model}

Lorentz-like models have already been applied to the study of inelastic gases, in order to show non-equipartition of energy for instance \cite{piasecki}. These systems are composed by {\em particles} that do not interact between themselves, and  undergo collisions with randomly 
distributed static (infinitely heavy) scatterers. In this paper, we will consider a model which is very similar to the Ehrenfest wind tree model \cite{ehrenfest}:
 
 \begin{equation}
 \frac{\partial f(v;t)  }{\partial t}   = |v| ~ (\frac{1}{\alpha^2} 
f(v^{'};t) - f(v;t))
\label{brenig1}
\end{equation} 
where the pre-collisional is defined by $v^{'} = - \frac{1}{\alpha} v$.
This kinetic equation has  been shown \cite{yves,lambi2} to be related to the inelastic Liouville equation  \cite{brenigR} and to population dynamics \cite{yves}. In the following, we use an IMM-like approximation in order to treat this kinetic equation, i.e. we assume that the collision rate $|v|$ may be replaced by a mean value $\sqrt{T(t)}$ and we rescale the time scale. We call the resulting model the Inelastic Lorentz Walk (ILW):

 \begin{equation}
 \frac{\partial f(v;t)  }{\partial t}   =  \frac{1}{\alpha} 
f(\frac{v}{\alpha};t) - f(v;t) ~~  \Leftrightarrow ~~ \frac{\partial \phi(k;t)   }{\partial t} =  \phi(\alpha k;t) - \phi(k;t)  
\label{ilw}
\end{equation} 
The right side equation is the kinetic equation for the characteristic function of the velocity distribution $\phi(k;t) = \frac{1}{2 \pi} \int dv f(v, t) e^{i k v}$.
 In Fourier space, the discrete time version of the process reads:
 
 \begin{equation}
  \phi_{N+1}(k) =  \phi_{N}(\alpha k) 
\label{discreteWalk0}
\end{equation} 
whose solution is:
\begin{equation}
  \phi_{N}(k) =  \phi_{0}(\alpha^N k)  
\label{discreteWalk}
\end{equation} 

 Let us stress that the stochastic process leaves the solution scale invariant, i.e. \ref{discreteWalk} is a self-similar solution whose form is that of the initial velocity distribution and whose temperature asymptotically vanishes.  Obviously, the discrete process does not lead to the formation of high energy tails, but this property ceases to be true if only a fraction p of the discs performs a collision at each step. In this case, the state of the system after one step writes:
\begin{equation}
 \phi_{1}(k) =  (1-p) ~ \phi_{0}(k)  +p ~ \phi_{0}(\alpha k)
\label{semiDiscreteWalk}
\end{equation} 
 Simple analytical calculations show that the kurtosis $\kappa \equiv \frac{<v^4>}{<v^2>^2}$ of the velocity distribution increases through \ref{semiDiscreteWalk} whatever the initial velocity distribution and proportion p$\neq1$. This is due to the fact that the superposition of identical distributions with different temperatures $T_1$, $T_2$ has a higher energy tail than the same distribution with the mean temperature $\overline{T}$$\equiv$(1-p) $T_1$+ p $T_2$  \cite{beck, ausloos1}.

Contrary to the non-linear IMM, the link between the discrete and the continuous time dynamics is straightforward. Indeed, usual methods of random walk theory \cite{balescu} lead to:

 \begin{equation}
 \phi(k;t) =  \sum_{N=0}^{\infty} \frac{t^N e^{-t}}{N!}  \phi_{0}(\alpha^N k)  
\label{continuousSol}
\end{equation} 
The system at time t is thus composed by  different classes of particles that are characterised by their mean energy $\alpha^N$, or equivalently by the number $N$ of undergone collisions. The proportion of particles in these classes is given by the probability for a particle of having performed $N$ collisions before time t, namely $\frac{t^N e^{-t}}{N!}$.
 On should note that \ref{continuousSol} is a superposition of the initial distribution at different energies, as in \ref{semiDiscreteWalk} and leads to the formation of high energy tails by the same mechanism. Assuming that the small k development of the initial characteristic function is
$
 \phi_0(k;t) =  \sum_{i=0}^{\infty} \frac{c_{0i} k^{i}}{i!} 
$, the formal solution \ref{continuousSol} reads:

 \begin{equation}
 \phi(k;t) =  \sum_{i=0}^{\infty}  \frac{c_{0i} k^{i}}{i!}  e^{-(1- \alpha^{i}) t}
\label{continuousSol2}
\end{equation}
Detailed analysis of this solution shows that the inelastic Lorentz walk exhibits multiscaling properties, i.e. the dimensionalized velocity moments $m_i \equiv (<(v^2)^{i}>)^{\frac{1}{2 i}}$ decay with different cooling rates $m_i \sim e^{-\lambda_i t}$. The cooling rates are given by $\lambda_i = \frac{1-\alpha^{i}}{i}$ and  are decreasing function of i: $\lambda_i<\lambda_j$, $i>j$. Consequently, the system is characterised by an infinite number of independent cooling rates. Moreover, the velocity moments $m_i$, $i\geq 2$, grow towards infinity as compared to $m_1$.

\section{Conclusions}

In this paper, we study the formation of the high energy tails observed in inelastic gases. To do so, we  focus on the one-dimensional Inelastic Maxwell Model, which is a mean field approximation of the Boltzmann equation for inelastic hard rods. 
By comparing a discrete and a continuous time version of the process, we show how  fluctuations of the number of  collisions imply the emergence of high energy tails.
To do so, we perform DSMC simulations of both dynamics, thereby  highlighting a new relation between number of {\em inelastic } collisions undergone by a particle and its average energy.
This relation is also  analytically studied by focusing on a simpler linear model, i.e. the  Inelastic Lorentz Walk. Let us stress that this mechanism is specific to inelastic gases, where energy is dissipated at each collision, and has no counterpart in elastic gases. 
 A generalisation of this work to higher dimensional systems and to more general kinetic models is under progress.

{\bf Acknowledgments}
R.L. would like to thank warmly M. Mareschal and M. Ausloos.  This work has been done thanks to the financial support of FRIA and ARC.


\begin{thebibliography}{0}

\bibitem{ehrenfest}
P. and T. Ehrenfest, {\em The conceptual foundations of the statistical approach in mechanics}, {Cornell Univ. Press} {(1959)}
 
\bibitem{bennaim1}  
E. Ben-Naim and P.L. Krapivsky, {\em Phys. Rev. E}, {\bf 61}{(2000)} {R5}


\bibitem{ernst}  
M.H. Ernst and R. Brito, {\em Europhys. Lett.}, {\bf 58} {(2002)} {182}  

\bibitem{ernst}
M.H. Ernst and R. Brito, {\em Phys. Rev. E}, {\bf 65} (2002) 040301(R) 

\bibitem{biben}  
 A Barrat,  E Trizac and M.H. Ernst, {\em J. Phys: Condens. Matter}, {\bf 17}  (2005) S2429

 \bibitem{barrat}
 A Barrat, T Biben, Z R\'acz, E Trizac and F van Wijland, {\em J. Phys. A}, {\bf 35} (2002) 463

\bibitem{balda1}  
A. Baldassarri, U.M.B. Marconi and A. Puglisi, {\em Europhys. Lett.}, {\bf 58} {(2002)} {14} 

\bibitem{lambi}  
R. Lambiotte and L. Brenig, {\em Phys. Lett. A}, {\bf 345} (2005) 309

 \bibitem{bird} 
G.A. Bird, {\em Molecular Gas Dynamics}, 
Clarendon Press (1976)

\bibitem{soto}  
R. Soto and M. Mareschal, {\em Phys. Rev. E}, {\bf 63} {(2001)} {041303}      

 \bibitem{wild}
  E. Wild,
  {\em Proc. Cambridge Philos. Soc.}, {\bf 47} {(1951)}, {602} 
      
\bibitem{yves}  
Y. Elskens, {\em J. Stat. Phys.}, {\bf 101} {(2000)} {397}  

\bibitem{lambi2}  
R. Lambiotte, PhD thesis, available at {\em http://users.skynet.be/lambiotte}

\bibitem{brenigR}  
J.M. Salazar and L. Brenig, {\em Phys. Rev. E}, {\bf 46} {(1999)} {2093}  
      
 \bibitem{balescu}
R. Balescu, {\em Statistical dynamics. Matter out of equilibrium}, {Imperial college press} {(1997)}
           
  \bibitem{beck}
  C. Beck, {\em Phys. Rev. Letters}, {\bf 87} {(2001)} {180601} 
  
  \bibitem{ausloos1}
  R. Lambiotte and M. Ausloos, cond-mat/0508773
  
  \bibitem{piasecki}  
P.A. Martin and J. Piasecki, {\em Europhys. Lett.} {\bf 46} {(1999)} {613}
  

 
  
  
\end{thebibliography}
\end{document}